\documentstyle{article}

\def\tr{{\rm tr}}

\def\half{{\textstyle {1 \over 2}}}
\def\quart{{\textstyle {1 \over 4}}}
\def\thh{{\textstyle {3 \over 2}}}

\def\prl{Phys. Rev. Lett.}
\def\cmp{Comm. Math. Phys.}
\def\quart{{\textstyle {1 \over 4}}}

\def\calV{{\cal V}}
\def\calP{{\cal P}}
\def\calS{{\cal S}}

\def\calD{{\cal D}}
\def\mh{{\hat \mu}}
\def\nh{{\hat \nu}}
\def\rh{{\hat \rho}}
\def\mm{{\hat m}}
\def\nn{{\hat n}}

\def\Rfour{t^8t^8 R^4}
\def\lam16{\lambda^{16}}

\def\tr{{\rm tr}}

\def\half{{\textstyle {1 \over 2}}}
\def\quart{{\textstyle {1 \over 4}}}

\def\thbar {\bar{\theta}}

\def\II{\relax{{\rm I}\kern-.10em {\rm I}}}

\def\ele{\relax{1\kern-.10em 1}}

\def\xxx#1 {{hep-th/#1}}

\def\npb#1(#2)#3 { Nucl. Phys. {\bf B#1} (#2) #3 } \def\rep#1(#2)#3
{ Phys.
Rept.{\bf #1} (#2) #3 } \def\plb#1(#2)#3{Phys. Lett. {\bf #1B} (#2) #3}
\def\prl#1(#2)#3{Phys. Rev. Lett.{\bf #1} (#2) #3}
\def\physrev#1(#2)#3{Phys. Rev. {\bf D#1} (#2) #3} \def\ap#1(#2)#3{Ann.
Phys. {\bf #1} (#2) #3} \def\rmp#1(#2)#3{Rev. Mod. Phys. {\bf #1}
(#2) #3}
\def\cmp#1(#2)#3{Comm. Math. Phys. {\bf #1} (#2) #3}
\def\mpl#1(#2)#3{Mod. Phys. Lett. {\bf #1} (#2) #3}
\def\ijmp#1(#2)#3{Int.
J. Mod. Phys. {\bf A#1} (#2) #3}

\newcommand{\be}{\begin{equation}}
\newcommand{\ee}{\end{equation}}
\newcommand{\bes}{\begin{eqnarray}\displaystyle}
\newcommand{\ees}{\end{eqnarray}}

\begin{document}

\baselineskip=14pt
\pagestyle{empty}
{\hfill DAMTP/97-109}
\vskip 0.1cm
{\hfill PUPT-1733}
\vskip 0.1cm
{\hfill hep-th/9710151}
\vskip 2cm
\centerline{SIXTEEN-FERMION AND RELATED TERMS IN M-THEORY
ON $T^2$}
\vskip 1cm
 \centerline{ Michael B.
Green$^{a,}$\footnote{m.b.green@damtp.cam.ac.uk},
Michael Gutperle$^{b,}$\footnote{gutperle@feynman.princeton.edu}
and Hwang-hyun
Kwon$^{a,}$\footnote{h.kwon@damtp.cam.ac.uk}}

\vskip 0.5cm
\centerline{$^a$DAMTP, Silver Street,}
\centerline{ Cambridge CB3 9EW,  UK}

\vskip 0.5cm
\centerline{$^b$Joseph Henry Laboratories,}
\centerline{Princeton University,}
\centerline{Princeton, New Jersey 08544,  USA}
\vskip 1.4cm
\centerline{ABSTRACT}
\vskip 0.3cm

Certain one-loop processes in eleven-dimensional supergravity
compactified
on $T^2$ determine exact,  non-perturbative, terms in  the
effective action
of type II
string theories compactified on a circle.  One example  is
the modular invariant   $U(1)$-violating interaction
of sixteen
complex spin-$\half$ fermions of  ten-dimensional  type IIB theory.
This  term,
together with  the (curvature)${}^4$ term, and many  other terms of
the same
dimension are all explicitly related by  supersymmetry.

\newpage

\pagestyle{plain}
\setcounter{page}{1}

The interconnections between different string theories and
eleven-dimensional supergravity provide strong constraints on the
possible form
of
various terms
in the low energy expansion of the effective action of M-theory and its
compactifications.  The
relationships  between   classical eleven-dimensional supergravity
\cite{cremmerscherk} and the   classical  type II supergravity
theories are
well known \cite{witten,aspinwall,schwarz}.   There is also some
understanding
of  certain special terms of  higher  order  in the low energy expansion.
For example,   absence of  five-brane anomalies and other arguments
determine
the presence
of an eleven-form, $C^{(3)} \wedge X_8$
\cite{vafawitten,duffminliu}  (where
$X_8$ is an eight-form made from the curvature).    Another
example is the
$\Rfour$ term\footnote{The
expression $\Rfour$ is used to indicate the particular contraction
of four
Riemann curvatures with the rank-eight tensor $t_8$ which is
defined in $SO(8)$
notation by $t^{ i_1 \cdots i_8} = \tr (\gamma^{i_1i_2} \gamma^{i_3i_4}
\gamma^{i_5i_6}
\gamma^{i_7i_8})$ where the $SO(8)$ vector indices, $i_r$ are extended to
$SO(10,1)$ indices in the eleven-dimensional theory.} that can be
deduced from
a one-loop quantum calculation in eleven-dimensional supergravity
\cite{ggv} .
  Compactifying on
 $R^9 \times T^2$ and identifying  the complex
structure of $T^2$   with the complex  scalar field, $\rho$, of  type IIB
superstring  theory leads to an exact determination of the
$\Rfour$ term in
the type IIB effective action, including
perturbative and non-perturbative terms.  This reinforces the
suggestion that
the    coefficient of the  exact $\Rfour$ term in the
effective  superstring action is a  specific non-holomorphic
modular function
of the
scalar fields, $f(\rho,\bar \rho)$  \cite{greengut,greenhove}.        One
implication of this suggestion  is that
there should be a non-renormalization theorem  for the $\Rfour$
term in type II
string
theory that
prevents perturbative
contributions beyond
one loop.   Recently,  very strong evidence for such a
non-renormalization
theorem  has been obtained \cite{anton,berko}.  There has also been
related
work on compactification to lower dimensions \cite{compacts} and
heterotic/type
I duality \cite{ threshold}.

To understand the structure of these terms recall  that in the IIB theory
\cite{nahm,greenschwarz,westschwarz,schwarz1,howewest} the complex field
 parameterizes the coset $SL(2,R)/U(1)$ (or, equivalently, $SU(1,1)
/U(1)$) where the $U(1)$ acts locally on the fermions in the theory.
The  $SL(2,R)$ symmetry of classical  IIB supergravity is broken in
string
theory (already at tree level) to $SL(2,Z)$   and
$\rho = \rho_1 + i \rho_2  = C^{(0)} + i
e^{-\phi }$ (where $\phi $
is the dilaton and  $C^{(0)}$ the  Ramond--Ramond scalar)  spans
the fundamental domain, $SL(2,Z)\backslash  SL(2,R)/U(1)$.  This
means that
the $U(1)$   \lq R-symmetry'
 is also   broken and interactions arise in which the
$U(1)$ charge is   violated by even integers.
The characteristic feature of those terms in the IIB effective action
that are
protected by non-renormalization theorems is that they are $F$-terms, or
integrals  over sixteen fermionic zero modes ---
half the
Grassmann  coordinates of the full type II superspace.

Many of these protected terms, such as the $\Rfour$ term,  conserve
the $U(1)$
charge.   Other examples are
the purely
bosonic terms considered  at one string loop in \cite{grossloane} such as
$\partial^2 G G^* R^2$   (where $G$ is a complex combination of the field
strengths of the  two antisymmetric tensors),
$\partial^4 (G G^*)^2$, and other  interactions involving the
complex scalar.
The derivatives here simply balance the dimensions ---  the precise index
structure will be apparent later.
There are
also terms coupling two fermions to two
bosons, such
as $\partial^3 \lambda   \psi^* G^* R$ (where $\lambda$ is the complex
spin-$\half$
fermion,  $\psi$ is the gravitino), and four-fermion terms such as
$\partial^6 (\lambda
\lambda^*)^2$.  We will see later that the full
non-perturbative expressions for these terms in the effective
action have a
very simple origin in a supersymmetric expression (some ot these
terms have
also been discussed in   \cite{partouche}).

The main point of this paper will be to generalize the arguments of
\cite{ggv}\   to include  terms in the effective action which, in
the   type
IIB language,
do not conserve   the $U(1)$ charge.
Any Feynman diagram in IIB supergravity naively preserves
the classical symmetry but such diagrams are hopelessly divergent.   A
consistent
regularization  should allow for $U(1)$ breaking   since
it is in any  case
broken by the effects of D-instantons.       String theory
does not
possess the continuous symmetry even at tree level.   The vertices
coupling
the massive string states  only respect the discrete symmetry and,
as will be
seen explicitly later, there are $SL(2,R)$-violating terms in the
tree-level
effective action that are crucial in ensuring the $SL(2,Z)$
invariance of the
full theory.  Thus,
string theory  regulates   the perturbative  loop divergences  in a
manner that
breaks the $U(1)$ symmetry to   a discrete subgroup.  The  class of
$U(1)$-violating processes that is protected from
perturbative
renormalization beyond one loop is related by supersymmetry to the
$\Rfour$
term as will be described at the end of this paper.  It  must
therefore be
possible, as in the $U(1)$-conserving examples, to obtain the complete
effective action for these terms by calculating  eleven-dimensional
 one-loop
amplitudes and using the correspondence between M-theory on $T^2$
and  IIB on
$S^1$.  The main new feature, which we will discuss in detail, is the
identification of the fermions in the two theories which was not
discussed in  \cite{aspinwall,schwarz}.

We will concentrate particularly on the  example of  the $\lambda^{16}$
interaction.  This  is the analogue of
the 't Hooft
fermion
vertex in  standard Yang-Mills instanton backgrounds
\cite{greengut}.  Since
$\lambda$ carries
$U(1)$-spin $\thh$ in the conventions used in \cite{schwarz1},  this term
violates 24 units of charge --- the presence of D-instanton effects
breaks the
$U(1)$ to a discrete subgroup.  A  striking difference  from the
examples of
terms that conserve the $U(1)$ charge  is that  the supergravity
one-loop
diagram  for this process {\it vanishes} in the eleven-dimensional
limit,  in
which the volume of the torus is infinite.   For finite volume the
 only
configurations that contribute are those  in  which the particle
circulating in
the loop winds around at least one cycle of the torus a non-zero
number of
times.  The zero winding number sector, which would have been
divergent, is
absent and the loop diagram is really finite.

The resulting $\lam16$ term has a very simple form with a coefficient,
$f_{16}$, that is a non-holomorphic   modular form of weight $(12,
-12)$ that
is
simply related to the $\Rfour$ term.    In this
notation  $(m^+, m^-)$  denotes the holomorphic and
anti-holomorphic weights and a function transforming with a $U(1)$
charge $2j$
has weight $(-j,j)$.   Furthermore, supersymmetry
relates
these
terms to all   other
terms of the
same dimension that are  integrals over half the superspace.   This
will be
expressed in a compact form   by an on-shell superspace  expression.

\medskip
\noindent{\bf One loop with four gravitons on $R^9 \times
T^2$}

We will begin by reviewing the   $\Rfour$ term
\cite{greengut}\  obtained in
\cite{ggv} by evaluating the four-graviton one-loop amplitude of
eleven-dimensional supergravity compactified  on a torus,  $T^2$, in the
directions $9$, $11$.     The volume of  the torus is    $\calV=
R_9 R_{11}$
and its complex
structure,  $\Omega
= \Omega_1 + i\Omega_2$, may be expressed in terms of the components
eleven-dimensional metric   $G_{\mh \nh}$ ( $\mh = 0, \cdots, 9,11$)  as
$\Omega_2 = G_{9\ 9} / G_{11 \ 11} =  R_9/R_{11}$ and  $\Omega_1 = G_{9\
11}/G_{11\ 11} = C^{(1)}$, where $C^{(1)}$ is the component of the
IIA one-form
along the direction $x^9$.
 The expression for the amplitude,   obtained rather efficiently  by
considering a first-quantized super-particle in the light-cone gauge
\cite{ggv},  has the structure,
\be
\label{loopgen}
A_{R^4}   =
\int d^9 p  {1
\over \calV} \sum_{m,n}\int \prod_{r=1}^4 d\tau_r  \tr(V^{(1)}_h(k^1)
V_h^{(2)}(k^2)
V_h^{(3)} (k^3)V_h^{(4)} (k^4) ),
\ee
where $V^{(r)}_h(k^r)$ is the graviton vertex operator for the $r$th
graviton with momentum $k^r$ and is
 evaluated at a proper time $\tau_r$ around the loop (and dimensional
quantities
are measured in units
of the eleven-dimensional Planck scale).   The  trace is
over the
fermionic zero modes and the loop momentum in the directions
$0,\cdots,8$
and  the  integers
$m$ and $n$
 label the Kaluza-Klein momenta in the directions $9$, $11$ of $T^2$.
The details
of the complete set of  light-cone vertices of eleven-dimensional
supergravity
are contained in \cite{ggk}  but will not be needed here.
  The trace over the fermionic modes gives an overall factor, $K$, which
contains eight powers of the external momenta and is the linearized
form of
$\Rfour$.    The $\Rfour$ term in (\ref{loopgen}) is  obtained by
setting the
momenta equal to zero in the remainder of the loop integral.  It is
convenient
to perform a double Poisson resummation in order to rewrite the sum
over the
Kaluza--Klein momenta  as a sum over windings, $\mm$ and $\nn$, of
the  loop
around the cycles of $T^2$, giving,
\bes
\label{loopgrav}
\calV  A_{R^4}  &=&   K \calV \sum_{\mm,\nn}\int {d \tau} \tau^{\half}
e^{- \tau \calV{ |\mm+\nn\Omega|^2\over \Omega_2}}  \nonumber\\
&=&  K\calV C+\pi^2  K\calV^{-\half} f(\Omega,\bar \Omega),
\ees
where
\be
\label{modfun}
f(\Omega,\bar{\Omega})=  \pi^{-2} \Gamma(3/2)
\sum_{(m,n)\neq(0,0)}{\Omega_2^{3/2}\over
|m+n\Omega|^3}
\ee
is a modular function.
The cubic ultraviolet divergence of $A_{R^4}$ is contained in the zero
winding term,   $\mm = 0 = \nn$,  which  has the divergent
coefficient, $C$.
It was argued in \cite{greenhove} that  in any regularization that is
consistent with
T-duality between the IIA and IIB string theories in nine
dimensions  this has
to be replaced by a regularized finite value, $C= \pi/3$ \cite{ggv}.
 Presumably a
microscopic description of M-theory (such as   matrix theory)
would reproduce
this value.    In the limit of zero volume, $\calV \to 0$, this term
disappears
and  only the term with coefficient $\calV^{-\half}$ survives.  After
appropriate rescaling of variables this reproduces the
$\Rfour$
terms in the effective action of ten-dimensional type IIB string
theory.  More
precisely, it reproduces the known tree-level and one-loop results
as well as
an infinite set of D-instanton corrections.  It remains a strong
conjecture
that there are no higher-loop corrections
\cite{greengut}.\footnote{However,
this
contradicts the claim in \cite{jengo} to have calculated a non-zero
coefficient
for this term at two loops --- clearly this should be investigated!}
The function $f$ in (\ref{modfun})  is the $s=\thh$  example of  a
generalized
(nonholomorphic) Eisenstein
series, or Maass  waveform \cite{tarras}, and is
an  $L^2$ eigenfuntion   of the  Laplace
operator on
the upper-half plane,
\be \label{laplaceeq}
\Omega_2^2\partial_\Omega \partial_{\bar{\Omega}}f =  {3\over 4}f.
\ee
This  Ward identity  follows  from the insertion of
two zero-momentum vertices for moduli fields in the four-graviton loop.

 \medskip
\noindent{\bf Fermions in  IIB on $S^1$ and M-theory on
$T^2$}

A key aspect of the duality symmetries is the connection between
M-theory on a
torus  of volume $\cal V$ and complex structure $\Omega$ and the type IIB
theory on a circle of circumference $r_B$ (in the string frame)
\cite{aspinwall,schwarz}.
   The
$SL(2,Z)$
duality of IIB string theory is thereby interpreted as the invariance
of M-theory on $T^2$ under the $SL(2,Z)$ group of large
diffeomorphisms of the
two torus.  The relation between the scalar fields of the two theories is
well-known to be given by
\be
\label{relat}
\Omega = \rho, \qquad r_B =  \calV^{-{3\over 4}} \Omega^{-\quart}_2 =
R_{9}^{-1} R_{11}^{-\half}.
\ee
The compactification radius in the IIB theory Einstein frame is  given by
$r^E_B = \calV^{-{3\over 4}}$.
The   three-form potential, $C^{\mh \nh\rh}$, is simply related to the
two-forms $B_N$ and $C^{(2)}$  of the IIB theory.   However,
 it will  also be important for us  to obtain the relationship between
the fermions
in the two theories.   To do this we begin by reviewing the
symmetries of  IIB supergravity.

(a)  {\bf Type IIB}

The scalar fields, which   parameterize the coset $SL(2,R)/U(1)$,
couple to the
 fermions  in a  locally $U(1)$ invariant manner. This coupling  is
described with the aid of the vielbein which is given (in  a
complex basis) by
the $SL(2,R)$ matrix,
\be
V= \left(\begin{array}{cc} V_+^1&V_-^1\\
V_+^2&V_-^2\end{array}\right) = {1\over \sqrt{2i
\rho_2}}\left(\begin{array}{cc}
  \rho e^{i\phi}& \bar{\rho}e^{-i\phi}\\
 e^{i\phi}& e^{-i\phi}\end{array}\right),
\ee

The group $SL(2,R)$ acts by matrix multiplication from the left and
the local
$U(1)$ acts from the right, so that under the combined transformations,
\be\label{vfixed}
V\to V' = {1\over \sqrt{2i \rho_2}} \left(\begin{array}{cc}
a&b\\
c&d
\end{array}\right)\left(\begin{array}{cc}
 \rho e^{i\phi}&  \bar{\rho}e^{-i\phi}\\
 e^{i\phi}& e^{-i\phi}\end{array}\right)\left(\begin{array}{cc}
 e^{i\alpha}& 0\\
 0&e^{-i\alpha}\end{array}\right),
\ee
where $ad-bc=1$.   The $U(1)$ acts only on the
fermions and on $V$ while the $SL(2,R)$ acts only on $V$
and the  second-rank   antisymmetric tensor potentials, $B^\alpha$ (where
$B^1=B_N$ is in the NS--NS sector   and  $B^2 = C^{(2)}$ in the
R--R sector).
We will
fix the   $U(1)$  gauge  by setting $\phi=0$,\footnote{The
gauge choice in \cite{schwarz1} was made in the $SU(1,1)$
parameterization and
corresponds to  a different   choice of gauge.}  so that a compensating
gauge transformation accompanies an $SL(2,R)$ transformation in
order to
maintain the gauge condition.
In this case a general  $SL(2,R)$ transformation transforms  the
complex scalar
by
\be
\rho \to {a\rho+b\over c\rho+d},
\ee
where  the compensating $U(1)$ transformation is given by
\be\label{phasealpha}
e^{i\alpha}=\left({c\bar \rho+d\over c \rho+d}\right)^{\half},
\ee
which induces  chiral transformations on the complex   spin-$\half$ and
spin-$\thh$ fermions,
 \be
\label{fermtrans}
\lambda \to  \left({c\bar \rho+d\over
c\rho+d}\right)^{3\over 4 }\lambda =  e^{\thh i \alpha \Gamma^{11}}
\lambda,\qquad
\psi_\mu \to  \left({c\bar \rho+d\over
c \rho + d}\right)^{\quart}\psi_\mu = e^{-\half  i\alpha  \Gamma^{11}}
\psi_\mu
\ee
($\mu = 0, 1, \cdots,9$)
which have $U(1)$ charges $\thh$ and $\half$, respectively and are
eigenstates
of $\Gamma^{11}$ of opposite chirality  (we have chosen
$\Gamma^{11}\lambda =
\lambda$, $\Gamma^{11} \psi_\mu = -\psi_\mu$).   Furthermore,
the complex  three-form field strength, $G=
\epsilon_{\alpha\beta}V^\alpha_+ d
B^\beta$  has $U(1)$  charge $1$  while the
 Maurer-Cartan form,
\be
\label{pdef}
P_\mu  =  -\epsilon_{ab} V_+^a \partial_\mu V_+^b=  {i\over 2}
{\partial_\mu
\rho\over
\rho_2},
\ee
has charge $2$.   The graviton $h_{\mu\nu}$ and the fourth-rank potential
$C^{(4)}$ are neutral under both  $U(1)$ and $SL(2,R)$.
The one-form,
\be
\label{qdef}
Q_\mu =  -i \epsilon_{ab} V_+^a\partial_\mu V_-^b= -{1\over 2}
{\partial_\mu
\rho_1\over
\rho_2},
\ee
 is a  composite $U(1)$  gauge connection.

Compactification of the IIB theory on a circle  $S^1$ of
circumference $r_B$
(in string frame)  in the
direction $x^9$
breaks the $SO(9,1)$ Lorentz symmetry to $SO(8,1)$.   The
complex chiral
spin-$\half$ fermion $\lambda$ simply becomes a  complex spinor of
$SO(8,1)$.  The gravitino
decomposes into a
nine-dimensional gravitino $\hat \psi_\alpha$ (where the $SO(8,1)$
vector index
$\alpha = 0, 1,\cdots, 8$) together with a second
complex  spin-$\half$ fermion,
\be
\label{newferm}
\chi^A = r_B\psi^A_9,
\ee
where the factor of $r_B$ comes from the component $e_{99}$ of the
zehnbein.
The nine-dimensional gravitino is defined by shifting $\psi_a$,
\be
\label{gravidef}
\hat \psi_a = \psi_a +  { 1\over 7} \Gamma_a \Gamma^9 \chi,
\ee
so that the kinetic term is diagonal.

\smallskip
(b) {\bf M-theory on $T^2$}

We will now identify the components of the $T^2$ compactification of
eleven-dimensional gravitino,  $\Psi_{\hat{\mu}}$   
($\hat{\mu}=0,\cdots,9,11$
),
 that correspond to the IIB fermions
on a circle.

Compactification  on $T^2$  breaks the local Lorentz symmetry
\ from $SO(10,1)$ to $SO(8,1)\times SO(2)$.   The
nine-dimensional fermions can then be organized into eigenstates of
$SO(2) \equiv U(1)$, which  is easily related to  the
$U(1)$  in
the denominator of  the coset space of the IIB theory.
The  world indices  split into the
compact   directions $\sigma =9,11$ and the noncompact directions
$\alpha = 0,\cdots, 8$.    We  will  make a block diagonal ansatz
for the
eleven-dimensional
elfbein, $e_{ \hat{\mu}}^{\ \hat{m}}$,
\be\label{block}
e_{\hat{\mu}}^{\ \hat{m}}=\left(\begin{array}{cc}
 e^{\ s}_{\sigma}  &0\\
0&  e^{\ a}_{\alpha}
 \end{array}\right),
\ee
where  $a$
labels the nine-dimensional tangent space ($a = 0,1 \cdots,8$)
and $s$ the
two-dimensional tangent space ($s=1,2$).
The zweibein, $e^{\ s}_{ \sigma}$,  of  $T^2$ may be chosen, in a
special Lorentz
frame, to be
 \be\label{zweibein}
e^{\ s}_{\sigma} =\sqrt{\calV\over  \Omega_2}\left(\begin{array}{cc}
\Omega_2 & \Omega_1\\
 0&1\end{array}\right).
\ee
Symplectic reparametrizations of the torus act as
$SL(2,R)$ matrices from the left and local  Lorentz
transformations
act
as $SO(2)$ transformations from the right. The  condition that
the zweibein
 remains  in the frame
(\ref{zweibein})
 leads to the standard $SL(2,Z)$ transformation of the complex
structure of the
torus,
\be
\Omega \to {a\Omega+b\over c\Omega+d}.
\ee
and induces a specific $\Omega$-dependent $U(1)$ transformation on  the
fermions.

We may write the spin-$\half$ components of the compactified
gravitino  in a
complex basis, $z=x_{11}+ ix_{9}$,  $\bar{z}=x_{11}
-ix_{9}$,   as   $\Psi_z = \Psi_{11} +  i \Psi_9$ and $\Psi_{\bar z} =
\Psi_{11} - i\Psi_9$.    To relate these 32-component spinors  to
$\lambda$ and
$\chi$ of the IIB theory we  shall first   organize them into
eigenstates of
the $U(1)$ rotations of  the compact $T^2$ generated by $i\Gamma^{11}
\Gamma^{9}/2
+ j_{vec}$, where $j_{vec}$ acts on the vector index so that
$j_{vec} \Psi_z
=\Psi_z$ and  $j_{vec} \Psi_{\bar z} = - \Psi_{\bar z}$.
Those components of the  spin connection   that have  tangent-space
indices in
the  compact directions are given by
\be
\omega_{z, \alpha z} =  {i\over 2}  { \partial_\alpha{\Omega} \over
\Omega_2},
\qquad
\omega_{ \bar{z}, \alpha\bar{z}} =  {i\over 2}  {
\partial_\alpha{\bar \Omega}
\over \Omega_2} ,\qquad
 \omega_{ z, \alpha\bar{z}}= \omega_{ \bar{z},\alpha z}= {1\over \calV}
\partial_\alpha
\calV.
 \ee
The IIB scalars are identified by  $P_\alpha = \omega_{z, \alpha
z}$, $\bar
P_\alpha = \omega_{ \bar{z}, \alpha\bar{z}}$ and  $\omega_{ z,
\alpha\bar{z}}
=- 4\partial_\alpha \ln r_B^E/3$ .

The $\Gamma$ matrices, defined in the complex basis by
 $\Gamma_z={1\over 2}(\Gamma^{11}+i\Gamma^9)$ and
$\Gamma_{\bar{z}}={1\over
2}(\Gamma^{11}-i\Gamma^9)$,   obey
\begin{equation}
\Gamma_z^2=\Gamma_{\bar{z}}^2=0, \qquad
\{\Gamma_z,\Gamma_{\bar{z}}\}= 1,
\end{equation}
so that the  combinations
 \be
\calP_z= (\calP_{\bar z})^T =\Gamma_{\bar{z}}\Gamma_{{z}}={1\over
2}(1-i\Gamma^9\Gamma^{11}),\qquad
\calP_{\bar{z}}= (\calP_z)^T = \Gamma_{z}\Gamma_{\bar{z}}={1\over
2}(1+i\Gamma^9\Gamma^{11})
\ee
(where the superscript $T$ indicates the transpose) are projectors,
satisfying
$\calP_z^2=\calP_z$, $\calP_{\bar z}^2 = \calP_{\bar z}$ and $\calP_z
\calP_{\bar z} =0$.

The action of these projectors on a 32-component real spinor  is to
replace it
with a complex sixteen component spinor  which is an eigenstate of
the $U(1)$
spin, $i \Gamma^{11}\Gamma^{9}/2$.   Therefore, we  may identify the four
fermions of
the nine-dimensional compactified theory with the components,
\bes\label{fermcoms}
&&\calP_z \Psi_z, \qquad \calP_{\bar z} \Psi_{\bar z}, \\
&&  \calP_z
\Psi_{\bar z}, \qquad  \calP_{\bar z} \Psi_z.\label{fermcoms2}
\ees
The components  (\ref{fermcoms}) have $U(1)$ charges $+\thh$ and
$-\thh$,   respectively,  while the components
(\ref{fermcoms2}) are a mixture of the states with  $U(1)$ charges
$\pm \half$.
These fields are simply related to the spin-$\half$ fermions of the
IIB theory
by converting $\lambda$ and $\chi$ (and their complex conjugates)  from
eigenstates of  $\Gamma^{11}$ to eigenstates of $i\Gamma^{11} \Gamma^9$.
This leads to the identifications,
 \be
\label{ferms}
  \calP_{z}\Psi_z=
\Gamma_{\bar{z}}\lambda , \qquad \calP_{\bar{z}}\Psi_z= \Gamma_z \chi
\ee
which gives  the eigenstates of $\Gamma^{11}$,
\be
\label{lamdeef}
\lambda = (1 + \Gamma_z) \calP_z \Psi_z, \qquad \chi  = (1 +
\Gamma_{\bar z})
\calP_{\bar z} \Psi_z.
\ee

In terms of the real components
\be\label{lamcoms}
\lambda_1=\psi_{11}^+-\Gamma^9 \psi_9^-, \qquad
\lambda_2=\psi_9^+   +\Gamma^9\psi_{11}^-.
\ee
where the superscripts ${}^\pm$ indicate the chirality (the value of
$\Gamma^{11}$).
Similarly, writing $\chi = \chi_1 +i \chi_2$ and multiplying
(\ref{ferms})  by
 $(1+\Gamma^{11})$  leads to
\be\label{chicoms}
\chi_1 = \Gamma^9 \psi_9^+ - \psi_{11}^-,\qquad  \chi_2 = -\psi_9^-
- \Gamma^9
\psi_{11}^+.
\ee

The remaining components,  $\Psi_\alpha$ form the  nine-dimensional
gravitino
after a shift similar to  (\ref{gravidef}),
\begin{equation}\label{mgravi}
\hat \Psi_a=  \Psi_a +  {1\over 7} \Gamma_a
\left(\Gamma_z\Psi_{\bar{z}}+\Gamma_{\bar{z}}\Psi_{z}\right)
 = \Psi_a +  {i\over 7} \Gamma_a \Gamma^9(\calP_z \chi - \calP_{\bar z}
\chi^*).
\end{equation}
The relations between the M-theory  fields and the IIB fields can
be confirmed
by  comparing the way they behave under supersymmetry
transformations.

For completeness we may also relate the IIB fermions to the IIA
fermions in the
obvious manner.
The spin-1/2 fermions in the IIA theory follow by direct
dimensional reduction
of eleven-dimensional supergravity \cite{campbellwest}.   The
nine-dimensional
fermions
that arise
\ from  the spin-$\half$ fields in ten dimensions are
\be
\lambda_{A}=\psi_{11}+\Gamma^9\Gamma^{11}\psi_9,
\ee
which decomposes into the chiral components,
\be\label{lamchir}
\lambda_{A}^+=\psi_{11}^+-\Gamma^9\psi_9^-, \qquad
\lambda_{A}^-=\psi_{11}^-+\Gamma^9\psi_9^+.
\ee
Similarly,
\be
\chi_{A}=\psi_9 +\Gamma^9\Gamma^{11}\psi_{11},
\ee
which  has   chiral components
\be\label{chichir}
\chi_{A}^-=\psi_9^- +\Gamma^9\psi_{11}^+, \qquad
\chi_{A}^+=\psi_9^+ -\Gamma^9\psi_{11}^-.
\ee
Comparing (\ref{lamcoms}) with (\ref{lamchir})  and  (\ref{chicoms}) with
(\ref{chichir})
gives the identification of fields of IIA and IIB in nine dimensions,
\be\label{fermrels}
\lambda_1 = \lambda_{A}^+, \qquad
\lambda_2 = \Gamma^9 \lambda_{A}^-,\qquad
\chi_2=-\chi_{A}^-,\qquad
\chi_1= \Gamma^9 \chi_{A}^+,
\ee
which are in agreement with the world sheet T-duality
rules.

\medskip
\noindent{\bf One loop with 16 $\lambda$'s on $T^2 \times
R^9$}

The  calculation of the loop amplitude is
facilitated by
using the light-cone superspace description of  the eleven-dimensional
super-particle,  as with the $\Rfour$ loop in \cite{ggv}.  However,  the
$\lambda^{16}$ term  of
interest  is given by a zero-momentum process which is simple to
calculate
without using the full   gravitino vertex operators    (they will
appear in
\cite{ggk}).    The zero-momentum loop
amplitude has the
structure (ignoring an overall constant)
\be \label{sixteenl}
A_{\lam16}   ={1\over \calV} \sum_{m,n}  \int d^9 p  \int
{d\tau\over \tau}
\tau^{16} \tr( V_\lambda^{(1)}
(0)\cdots V_\lambda^{(16)} (0))   e^{ - \tau (p^2+{1 \over \calV\Omega_2}
|m+n\Omega|^2)},
\ee
 where  $V_\lambda^{(r)}(0)$ is the zero-momentum vertex for the
$\lambda$
component of the $r$th gravitino.    This is independent of the
proper time and
the factor of $\tau^{16}$ is simply  the volume of integration over
the proper
times of the vertices around the loop.   The trace in the above
expression is
over the fermionic operators in the vertices.

The zero-momentum gravitino vertex is very simply expressed in
terms of the
eleven-dimensional supercharge, $Q^A$, by
\begin{equation}\label{gravivertss}
V_\Psi = \bar\zeta_\mh  Q  p^\mh,
\end{equation}
where $\zeta_\mh^A$ is the zero-momentum wave function, $p_\mh$ is
the momentum
of the particle circulating around the
loop and
$\bar\zeta_\mh \equiv \zeta_\mh \Gamma^0$ (with no complex
conjugation).    The expression (\ref{gravivertss}) can be derived
by imposing
the
conditions
that the $\Psi$, $h$ and $C^{(3)}$ vertices transform
into each other
under supersymmetry in the appropriate manner.  This  is the same
argument
conventionally used to derive superstring vertices.  It is most
explicit in
the light-cone formalism in which the  $SO(10,1)$ symmetry  is
 broken
to $SO(9) \times SO(1,1)$ before compactification (and to $SO(7)
\times SO(2)
\times SO(1,1)$ after) .  The supercharge  $Q^A$ is represented  on the
momentum  and a fermionic  $SO(9)$ spinor,  $\calS$,  satisfying
$\Gamma^+
\calS=0$ (where $\Gamma^+ = (\Gamma^0 + \Gamma^1)/\sqrt 2$).   As usual,
$\calS$ satisfies $\{\calS^A, \calS^B\} = (\Gamma^+\Gamma^-)^{AB}$.
 In writing
(\ref{gravivertss}) we have ignored  contact terms which do not
contribute to
the spin-$\half$ process that we are interested in below.

We now want to consider the components of $\Psi_\mh$ containing
$\lambda$.  Inverting  (\ref{lamdeef})  gives,
\begin{equation}\label{psilam}
\Psi_z = \calP_{\bar z} \Gamma_z \chi + \calP_z \Gamma_{\bar z}
\lambda, \qquad
\Psi_{\bar z} = \calP_z \Gamma_{\bar z}  \chi^* +  \calP_{\bar z}
\Gamma_z
\lambda^*.
\end{equation}
\ from which it follows that
\begin{equation}\label{newlam}
V_\lambda=-\bar \lambda \Gamma_{\bar z}  \calP_{\bar z}  Q p_{\bar
z} = -\bar
\lambda\Gamma_{\bar z}   q  p_{\bar z}  ,
\end{equation}
where $q =  \calP_{\bar z}  Q$ is a projected supercharge that
satisfies the
anticommutation relations
\begin{equation}\label{commsq}
\{q^A, q^B\} = \calP^{AC}_{\bar z} \calP^{BD}_{\bar z} \{Q^C, Q^D\} =
\Gamma_z^{AB} p_{\bar z}.
\end{equation}
The momentum dependence can be scaled out  by changing to   $\hat
q^A = q_A
(p_{\bar z})^{-\half} $, which satisfies $\{\hat q^A, \hat q^B\} =
\Gamma^{AB}_z$.  In the chirally projected subspace  $\Gamma^{AB}_z
\propto
\delta^{AB}$ and the commutation
relation does
not depend on $\bar z$.
In the  light-cone gauge formalism the vertex (\ref{newlam})
reduces  to the
components,
\be\label{vpsi2}
V_\lambda =   \sqrt{p^+}  \lambda\  \calS\   p_{\bar z} , \qquad
V_{\tilde{\lambda}}    = {1\over
\sqrt{p^+}}   {\tilde \lambda}\  \Gamma^9  \calS\   p_{\bar z}  
p_{\bar z},
\ee
where $\lambda$ and $\tilde \lambda$ here refer to the
$\Gamma^+\Gamma^-$ and
$\Gamma^- \Gamma^+$
projections on
the covariant spinor wave function (and $\Gamma^{9 AB} \equiv
\delta^{AB}$ in
the projected subspace).

Substituting the expression for the vertices (\ref{newlam}) into
(\ref{sixteenl})
and integrating over   the loop momentum gives an expression of the form
\be\label{resf}
\calV A_{\lam16}  = \hat K \calV^{-\half} f_{16} (\Omega, \bar \Omega),
 \ee
where the kinematic prefactor,
\be
\label{kinems}
\hat K = \bar\lambda^{(1)}_{A_1} \cdots \bar
\lambda^{(16)}_{A_{16}}  \tr (
\hat q^{A_1} \cdots \hat q^{A_{16}}),
\ee
is manifestly antisymmetric under permutations of the (commuting)
fermion wave functions  due to
the anticyclic property of the trace.  It  can be rewritten (up to
an overall
constant) as
\be
\label{kinfac}
\hat K \sim \bar\lambda^{(r_1)}  \Gamma^{\mu_1 \mu_2 \mu_3}
\lambda^{(r_2)}
\cdots \bar\lambda^{(r_{15})}  \Gamma^{\mu_{22} \mu_{23} \mu_{24}}
\lambda^{(r_{16})} T_{\mu_1  \cdots \mu_{24}}\epsilon_{r_1 \cdots
r_{16}},
\ee
where the antisymmetric tensor $T$  (which is analogous to the
$SO(8)$ tensor
$t^{i_1 \cdots i_8}$ defined earlier) may be defined in terms of a
Grassmann
integral,
\be
\label{pfaffe}
T^{\mu_1  \cdots \mu_{24}}  M_{\mu_1 \mu_2 \mu_3} \cdots
M_{\mu_{22} \mu_{23}
\mu_{24}}  = \int d^{16} \eta e^{M_{\mu\nu\rho}\bar \eta
\Gamma^{\mu\nu\rho}
\eta} ,
\ee
where $\eta$ is a  Grassmann variable.  The kinematic factor is $SO(9,1)$
covariant.  The rest of the expression (\ref{resf})  depends on the
circumference $r_B$ through the factor $\calV^{-\half}$.  The
$\Omega$ and
$\bar \Omega$ dependence  comes from the  loop integration and is
given by
\bes
\label{loopam}
f_{16} (\Omega, \bar \Omega)  &=&  \sum_{m,n} \int {dt\over t} t^{23/2}
 \left({1\over \sqrt{\Omega_2}} (m+n\bar \Omega)\right)^{24}
\exp\left(-t
 {1\over {\Omega_2}}|m+n\Omega|^2\right)\\
 &=& {1\over \sqrt{\Omega_2}}\Gamma (23/ 2)\sum_{m,n}
{ (m+n\bar \Omega)^{24} \over | m+n \Omega|^{23}}.    
\label{resultintegrals}
\ees
The convergence  properties of this expression will be discussed below.

The complete expression for the $\lambda^{16}$  field theory
interaction is
given (up to an overall constant) by
\be
\label{fields}
S_{\lam16} = \calV^{-\half} \int d^{16} \eta  e^{\bar \lambda
\Gamma^{\mu\nu\rho}\lambda \bar \eta \Gamma^{\mu\nu\rho}  \eta}
f_{16} (\Omega, \bar
\Omega) \sim  \calV^{-\half} \int d^{16} \eta  e^{\bar \lambda  \eta}
f_{16} (\Omega, \bar \Omega),
\ee
where $\lambda^A$ is here  the anticommuting  spin-$\half$ field. The
overall constant  can be fixed  by linearized supersymmetry in  the
manner described in the final part of this paper.

\medskip
\noindent{\bf Modular invariance }

The final expression (\ref{resultintegrals}) is superficially cubically
divergent due to the sum over the Kaluza-Klein charges.    However,
this does not take into account the phase dependence which, for
generic $\Omega$, can lead to a cancellation between the growing terms
in the sum. Indeed,
for asymptotically large $m$ and $n$ the sum over
discrete
momenta $p_z$ and $p_{\bar z}$ can be replaced by integrals and the
result is
proportional to
\be\label{contin}
\int dp_z dp_{\bar z} p_{\bar z}^{24} (p_z p_{\bar z})^{-23/2},
\ee
which vanishes if $|p_z|$ is regularized. Equivalently,  it  
vanishes if the
phase
integration is carried out before the integration
over $|p_z|$ which is the prescription that will be used in the  
following.

In order to extract the finite result it is useful, as before,  to
use a double
Poisson summation to transform from the discrete momentum sum to a
sum over
windings of the loop  around the torus.  Writing
\be
\label{doubpois}
 f_{16}  ={1\over \sqrt{\Omega_2}} \left({\partial \over \partial
\alpha}\right)^{24}
\sum_{m,n} \int {dt \over t} t^{23/2}\left.
 e^{-t|m+ n\Omega|^2 + \alpha (m+ n\bar \Omega)}\right|_{\alpha =0},
\ee
the Poisson resummation equates this to
\bes\label{finalfin}
f_{16}  &=&{1\over \sqrt{ \Omega_2}} \left({\partial \over \partial
\alpha}\right)^{24}
  \sum_{\mm,\nn}{\pi \over
\Omega_2}\int {dt \over t} t^{21/2} \left.  \exp\left( -{\pi^2 \over
t \Omega_2^2} |\nn + \mm \Omega|^2 + {i\pi \alpha \over t
\Omega_2}(\nn  + \mm\bar \Omega)
\right)\right|_{\alpha=0}\nonumber\\
&=&  {1 \over \pi^2 }   \Omega_2^{3/2}  \Gamma (27/ 2)
         \sum_{(\mm,\nn)\ne (0,0)} {(\nn + \mm\bar \Omega)^{24} \over
|\nn + \mm \Omega|^{27}},\label{Iresult}
\ees
where $\mm$ and $\nn$ are winding numbers.  Although the treatment
of the zero
winding-number term, $\mm = \nn=0$, looks somewhat ambiguous, the earlier
argument concerning the absence of an ultraviolet divergence
reinforces its
exclusion from the sum.

It is easy to see that under  $SL(2,Z)$  transformations from $\Omega$ to
$\Omega'$,
\be
\label{imod}
f_{16} (\Omega, \bar \Omega) \to f_{16} '(\Omega', \bar \Omega') =
\left({c\Omega+d\over
c\bar{\Omega}+d}\right)^{12}   f_{16} (\Omega, \bar \Omega)
\ee
which cancels the transformation of  $\lam16$  (\ref{fermtrans})
so that the
amplitude is invariant.

After translating from the M-theory coordinates ($\Omega$, $\calV$)
 to the string-frame  IIB
coordinates  ($\rho$, $r_B$) using (\ref{relat})  the  expression
$A_{\lam16}$
gives  a finite
contribution to the IIB   effective action in the limit, $r_B\to
\infty$, of
ten decompactified dimensions,
\be
\label{beffamp}
S_{\lam16} = \int d^{10} x\,  \det e\, \hat  K\,  f_{16}(\rho, \bar
\rho).
\ee
In writing
this string-frame expression   the
fermion field has been rescaled by $\lambda \to \lambda e^{\phi/8}$ in
transforming from  the Einstein
frame.
\medskip

\noindent
{\bf Large $\rho_2$ expansion}

It is of interest to expand (\ref{beffamp})   in the weak-coupling limit,
$\rho_2 =
e^{-\phi}\to \infty$,  in order to make contact with the terms that
arise in
type IIB perturbative string theory.    The leading term comes from
the terms
with $\mm=0$
in $f_{16} $,
\be
f_{16}^{\mm=0,\nn\ne 0} =   \pi^{-2}\Gamma(27/2)\zeta(3)
(\rho_2)^{\thh} ,
\ee
which has a $\rho_2$ dependence that  corresponds  to a string tree-level
contribution.
  Although we have not explicitly calculated this term in
string theory, it  must arise from the tree with sixteen $\lambda$ states
attached in much the same way as the $\Rfour$ term was found as a
nonleading
tree effect  \cite{grisaru,grosswitten}.

In order to take the large $\rho_2$ limit in the remainder of the terms
($\mm\ne 0$) it is necessary to do undo the  Poisson resummation on
the index
$\nn$, giving,
\bes\label{onepois}
&&f_{16}^{\mm \ne 0, \nn}  =   \pi^{-2} \rho_2^{\thh}
\Gamma(27/2)\sum_{\mm\neq 0,\nn}
{(\nn + \mm\bar{\rho})^{24}
\over |\nn + \mm \rho|^{27}} = {1 \over \pi^2} \rho_2^{\thh}
\sum_{\mm\neq 0,\nn} \int {dt\over t} t^{27\over 2}  (\nn +
\mm\bar{\rho})^{24}
\exp(-
 t|\nn + \mm \rho|^2)\nonumber\\
&=& {1 \over \pi^2} \rho_2^{\thh}
\sqrt{\pi}\left. \left({\partial\over \partial
\alpha}\right)^{24}\sum_{\mm\neq 0,n}\int
{dt\over t}
  t^{13}\exp\left( -{1\over t}(\pi n -\half i\alpha)^2 - t
\mm^2\rho_2^2 - 2\pi
i \mm n\rho_1 -i\mm \alpha \rho_2\right)\right|_{\alpha=0}.
\ees
The $n=0$ term in this expression is
\bes\label{loopcon}
f_{16}^{\mm\ne 0,n=0} & =& \pi^{-\thh}  \rho_2^{\thh}   \left.
({\partial\over
\partial \alpha})^{24}\sum_{\mm\neq 0}\int {dt\over t}
  t^{13}\exp\left( {1\over 4t}\alpha^2 - t \mm^2\rho_2^2  - i\mm \alpha
\rho_2\right)\right|_{\alpha=0}\nonumber\\
& = &  \pi^{-2} \Gamma(23/2)  \zeta(2)\rho_2^{-\half},
\label{nzero}\ees
which has the $\rho_2$-dependence characteristic of a one-loop
contribution.
It is  easy to see that the one-loop type IIB string amplitude with
sixteen external
$\lambda$'s can, indeed, be non-zero.   This is an intrinsically stringy
effect, in which massive string states circulate in the loop.
Again, we
have not checked  its coefficient with (\ref{loopcon}).

As with the $\Rfour$ terms there are no higher-loop contributions in
(\ref{onepois}).
 The remaining terms  which have   $\mm\neq 0$, $n\neq 0$ give
nonperturbative
contributions that are associated with D-instanton effects.  They can be
evaluated by making  use of the integral representation for  Bessel
functions,
\be
\int {dt\over t} t^{13} e^{-t - {z^2\over 4t}}= 2^{-12} z^{13}
K_{-13}(z)
\ee
and the asymptotic expansion
\be
K_{-13}(z)= \sqrt{\pi\over  2 z} e^{-z} \sum_{k=0}^\infty  {1\over
(2z)^k}{\Gamma(-13+k+\half)\over k! \Gamma(-13- k+\half)}.
\ee
The result is that the non-perturbative terms have the form,
\be
\label{nonperty}
f_{16}^{\mm \ne 0,  n\ne 0}= 2^{24} \pi^{23/2}\rho_2^{12} \sum_{\mm,n>0}
n^{25/2}
\mm^{21/2} e^{2\pi i \mm n \rho }(1 + O(\rho_2^{-1}))
\ee
So we see that only  instantons (and not anti-instantons)
contribute to each
power of  $e^{-\mm n \rho_2}$  at leading  order in $(\rho_2)^{-1}$.  The
leading contribution is therefore of the form $\rho_2^{12}$ times a
holomorphic
function of $\rho$.  The anti-instanton contribution starts at
$O(\rho_2^0)$.

The complete expression for the amplitude is obtained by adding the
tree-level,
one-loop and non-perturbative terms,
\be
\label{fulla}
f_{16}  = f_{16}^{\mm=0, \nn} + f_{16}^{\mm\ne 0, n=0} +
f_{16}^{\mm\ne 0,  n\ne 0}.
\ee
The structure of this expansion suggests that the $\lambda^{16}$
term in the
IIB string theory  does not get perturbative renormalizations
beyond one loop,
as in the case of the $\Rfour$ term (as should be evident from an
extension of
the arguments in \cite{berko}).

The relationship of  the $f_{16}(\rho,\bar \rho) \lam16$ term to the
$f(\rho,\bar\rho)\Rfour$ term is
illuminated by
introducing  the covariant derivative,    $\calD_d$, that  maps
modular  forms of
weight   $(d, \bar d)$ to forms of weight $(d+2, \bar d)$, and
its complex
conjugate,   $\bar \calD_d$, that maps modular  forms of
weight   $(d, \bar d)$ to forms of weight $(d, \bar d +2)$,
\be
\label{covder}
F_{d+2, \bar d} = \calD_d F_{d,\bar d} =  i \left( { \partial\over
\partial
\rho}  + {d
\over (
\rho - \bar \rho)}\right) F_{d,\bar d}, \qquad F_{d, \bar d+2} =
\bar \calD_d
F_{d,\bar d} =  - i \left( { \partial\over \partial \bar  \rho}  -  {d
\over (
\rho - \bar \rho)}\right) F_{d,\bar d}
\ee
where $F_{d,\bar d}$ is an arbitrary modular form of weight
$({d,\bar d})$ .
Using the notation,
\be
\label{seqd}
\calD^k F_{d,\bar d} = \calD_{d + 2(k-1)} \calD_{d+ 2(k-2)} \cdots
\calD_d
F_{d,\bar d},
\ee
it follows very simply that
the functions  $f_{16}$ and $f$ are related by
\be
\label{covrel}
f_{16} (\rho,\bar \rho)  =     \rho_2^{12} \calD^{12}f(\rho,\bar \rho).
\ee
and the effective action (\ref{beffamp}) can be written as
 \be
\label{beffact}
S_{\lam16} = \int d^{10} x  \det e\, \hat  K\, \rho_2^{12}\,\calD^{12}
f(\rho,\bar \rho) ,
\ee
up to an overall constant that will be fixed by supersymmetry.

\medskip
\noindent{\bf Linearized supersymmetry and other terms of the same
dimension}

The close connection between     $\Rfour$ and $\lam16$ terms is
obviously a
consequence of their  common
superspace origin which can be seen already in the rigid limit of the
linearized theory.   Consider a  linear  superfield
$\Phi(x,\theta)$  (where
$\theta$ is a complex chiral $SO(9,1)$ Grassmann spinor)  that
satisfies the
holomorphic constraint $ D^* \Phi=0$ and the on-shell condition
$D^4    \Phi
=  D^{*4} \Phi^*$ \cite{howewest} where
\be
\label{covderiv}
 D_A = {\partial \over \partial \theta^A} +2i (\Gamma^\mu  \theta^*)_A
\partial_\mu, \qquad   D^*_A = - {\partial\over \partial\theta^{*A}}
\ee
(recall that $\bar \theta = \theta\Gamma^0 $ does {\it not}
involve complex
conjugation) are the holomorphic and anti-holomorphic covariant
derivatives that
anticommute with
the rigid supersymmetries
\be
\label{susys}
Q_A ={\partial \over \partial \theta^A},
\qquad   Q_A^* = - {\partial \over \partial
\theta^{*A}} +  2i
(\bar \theta \Gamma^\mu  )_A \partial_\mu .
\ee
The field $\Phi$ has an expansion in powers of $\theta$ (but not $
\theta^*$), describing the 256 fields in an on-shell supermultiplet,
\bes
\label{phidef}
\Phi  &=& \rho_0 + \Delta\nonumber \\
 &= &\rho_0 + a  - {2i \over g}\thbar\lambda
  - {1 \over 24 g} \thbar\Gamma^{\mu\nu\sigma}\theta G_{\mu\nu\sigma}
+{i\over  6 g}\thbar\Gamma^{\mu\nu\sigma}\theta
\bar\theta\Gamma_{\nu}\partial_{\sigma}\psi_{\mu}  - {i \over 48
g}\thbar\Gamma^{\mu\nu\eta}\theta\thbar\Gamma_{\eta}^{\ \sigma\tau}\theta
R_{\mu\nu\sigma\tau}+
\cdots,
\ees
where $\Delta$ is the linearized fluctuation
around   a flat background with a constant scalar, $\rho_0 = \rho -  a =
C^{(0)}_0 +i g^{-1}$.   The coefficients of the component fields
are consistent
with the conventions used in \cite{schwarz1}.  The terms  indicated
by $\cdots$
 fill in the remaining members of the ten-dimensional $N=2$ chiral
supermultiplet, comprising (in symbolic notation)   $\partial dC^{(4)}$,
$\partial
\psi_{\mu\nu}^*$,  $ \partial ^2 G^*_{\mu\nu\sigma}$,  $\partial
^3\lambda^*$
and $ \partial^4\rho^*$.

Although the linearized theory cannot capture the full structure of
the terms
in
the effective action it can be used to relate   various terms in
the limit of
weak coupling, ${\rm Im} \rho_0 = g^{-1}\to \infty$ (where $g
=e^{\phi_0}$ is the
coupling constant).  The linearized
supersymmetric on-shell action  has the form
\be\label{actdef}
S' =   {\rm Re } \int d^{10}x  d^{16} \theta \,  g^4  F[\Phi],
\ee
which is manifestly invariant under the rigid supersymmetry
transformations,
(\ref{susys}).
The various  component interactions contained in (\ref{actdef})
are obtained
\ from the $\theta^{16}$ term in the expansion,
\be
\label{expands}
F[\Phi] = F(\rho_0) + \Delta {\partial \over \partial \rho_0}
F(\rho_o) + \half
 \Delta^2 \left( {\partial \over \partial \rho_0}\right)^2  F(\rho_o)   +
\cdots.
\ee

 The covariant derivatives  in the
  complete nonlinear relation  between the $\Rfour$ and $\lam16$
coefficients
(\ref{covrel})  differ from the  linear  derivatives in (\ref{expands})
because they contain inhomogeneous terms   $ \rho_2^{-1} d=   g d$.
 It is
clear that even in the $g\to 0$ limit such terms may be  as
important  as the
$\partial/\partial \rho_0$ term  when  $D$ acts on a function of  $g$.
However, the inhomogeneous term may be neglected and the  covariant
derivative
linearizes in the limit  $g\to 0$  when acting on the factors
$e^{2\pi i \mm n
\rho_0}$ of  the  instanton  terms.    Therefore, a linearized superspace
expression such as  (\ref{actdef}) must contain the exact  leading
multi-instanton contributions to the $\lam16$ and  $\Rfour$  terms -- the
$g^{-12}$  terms in  (\ref{nonperty}) and the corresponding $g^0$
non-perturbative terms in $f$.   This linearized expression cannot,
however,
contain the correct tree or one-loop terms, or the non-leading
contributions to
the instanton and  anti-instanton terms.  The  leading  instanton
terms arise
by  substituting  the expression
$F_{\mm n} =  c(\mm,n) e^{2\pi i |\mm n| \Phi}$
into  (\ref{actdef}) and the
$\lambda^{16}$ term in the expansion  (\ref{expands}) comes from the
$\Delta^{16}$ term.  Comparing this  with
(\ref{nonperty}) determines
\be
\label{coeffsdef}
 c(m,n) =  \mm^{-11/2} n^{-7/2}.
\ee
The single instanton term ($\mm =  n =1$) is the analogue of the
't Hooft
vertex in Yang-Mills theory and its form can be deduced directly by
semi-classical
quantization  of fermionic zero modes in the background of a D-instanton
solution of the type IIB supergravity.
The $\Rfour$ interaction  in   (\ref{expands})  comes from the
$\Delta^4$ term.
  Using the coefficients
(\ref{coeffsdef}) gives agreement with  the expression for the instanton
contributions in
\cite{greengut}
in the $e^\phi \to 0 $ limit.
After taking into account some combinatoric factors the relative
coefficients of  the $\Rfour$, and $\lam16$  terms is
\be
\label{pertrel}
f_{16}^p =  g^{-12}\left( {\partial \over \partial
\rho_0}\right)^{12}
f^p,
\ee
where $p= \mm n >0$ is the instanton number.  This coincides, as it  
should,
with  the large $\rho_2$
linearization  of
(\ref{covrel}).

All of the  $\theta^{16}$ terms that contribute to  the effective action
(\ref{actdef})  are
terms
of the same
dimension that  violate the $U(1)$ charge by $2n$ units.
 These include
the terms
mentioned in    the introduction  that  conserve the $U(1)$ charge  
($n=0$)
and  the
$\lam16$ term which violates the $U(1)$ charge by $24$ units   
($n=12$).  Among
the
 many other
contributions are terms, such as $\partial^4 (\psi^*\psi^*
\lambda\lambda)$ which violate the $U(1)$ charge by  the minimal non-zero
amount ($n=2$).

The covariantization of the derivatives  in the exact expression,
necessary for
$SL(2,Z)$ invariance,   must involve the full nonlinear
structure of the  theory.     This requires an infinite number of
   corrections
of higher order in $g$ to each instanton contribution.  It  is
clear  how such
terms arise in the stringy description of a single D-instanton
\cite{greengut}.  There, the leading effect arises from sixteen
disks with
Dirichlet boundary conditions in all directions with one
$\lambda$
and one
open-string fermion vertex operator attached to each disk.  The
nonleading
effects arise   from summing  diagrams  with less than sixteen
disks with any
subset of the sixteen $\lambda$'s   attached to each.    Such
 a simple
description does not apply to the multiply-charged D-instantons.
However,
consistency with $SL(2,Z)$ invariance suggests that the exact
 form
of the terms
in the effective action is given simply by replacing the
 ordinary
derivatives
in the expansion (\ref{expands})  by  covariant derivatives.

\vspace{.3cm}

{\bf Acknowledgements:}

\vspace{.1cm}

\noindent The work of  M. Gutperle is partially supported by a DOE grant
 DE-FG02-91ER40671, NSF grant PHY-9157482 and a James S.
 McDonnell grant 91-48. The work of H. Kwon is partially supported
 by a Dong-yung Scholarship.

\end{document}